# Studies of the Modular COsmic Ray Detector (MCORD) using an automatic temperature control loop to maintain constant gain parameters of semiconductor SiPM photomultipliers.

M. Bielewicz[1], M. Kiecana[1], A. Bancer[1,2], J. Grzyb[1], M. Grodzicka-Kobylka[1],
T. Szczesniak[1], K. Kopanski[2], W. Noga[2], L. Kazmierczak[1], G. Saworska[1], A. Broslawski[1],
P. Mazerewicz[1], E. Jaworska[1]

1. National Centre for Nuclear Research, Otwock-Swierk, Poland
2. The Henryk Niewodniczański Institute of Nuclear Physics Polish Academy of Sciences, Cracow, Poland

*Abstract*

The MCORD detector is a modular scintillator-based system employing silicon photomultipliers (SiPMs) and FPGA-based digital signal processing, designed for applications such as cosmic muon detection, veto systems, and detector calibration support. In this work, we investigate the influence of ambient temperature variations on detector performance, with particular emphasis on SiPM gain stability. Several automatic temperature compensation loops were implemented to stabilize the operating voltage of the sensors. Based on controlled laboratory measurements, we evaluate the effectiveness of different control strategies, including variations in temperature averaging time and threshold response criteria. The performance of each approach is compared in terms of gain stability and response dynamics. We identify the optimal temperature control configuration for planned MCORD measurements and present recent modifications to the detector electronics, including updated software for AFE control. Additionally, we describe modifications made to the detector's electronics since the previous publication, including new software developed to control the AFE electronics.

## 1. Introduction

Work on the Modular Cosmic Ray Detector (MCORD) began around 2017 [1] as part of the development of a detection system for the MPD detector in the NICA (Nuclotron‑based Ion Collider fAcility) [2]. During the initial phase, the detector concept was developed, a demonstrator was constructed, and a Conceptual Design Report (CDR) and Technical Design Report were prepared, forming the basis of the publication Conceptual Design Report of the MPD Cosmic Ray Detector (MCORD) [3]. In subsequent years, the team from the National Centre for Nuclear Research (NCBJ) continued the detector development, performing detailed analyses and design optimizations. Currently, the detector is not dedicated to any specific experiment, making it a versatile instrument for a range of applications.

The MCORD detector is based on plastic scintillators produced by NUVIA [4], with 2 mm-diameter optical fibers from Kuraray [5], and light is read out from both ends of each tile using silicon photomultipliers (SiPMs) from Hamamatsu [6]. Each tile is equipped with its own Analog Front End (AFE) electronics, and each SiPM sensor includes a preamplifier and a temperature sensor. Signals from eight tiles are collected in a HUB system, which transmits them to FPGA-based digital electronics without modifying the analog signal. A detailed analysis and calibration of both the AFE and digital electronics has been described in a review publication on the MCORD detector [7], covering ADC calibration methods, electronics stability, and the time required for the system to reach full stability after startup. During these studies, it was found that at very low currents the digital current meter exhibited significant noise (20–30 bits), affecting measurement reproducibility. The issue was diagnosed, and the AFE electronics were modified to minimize this effect – details of this correction are presented in Chapter 5 (MCORD Electronics Correction).

The current focus of the work is on the temperature dependence of MCORD operation. Due to the high sensitivity of SiPM gain to temperature variations [8], a Temperature Loop (TL) function was added to the AFE software. Under stable laboratory conditions, this effect is negligible; however, in variable environments, a supply voltage correction is required to compensate for temperature-induced gain changes. The TL loop measures the temperature near each SiPM sensor, averages the readings, and adjusts the sensor supply voltage based on a predetermined correction factor. The method for determining this correction factor is described in Chapter 3 (SiPM Coefficient Factor Measurement).

In this paper, we describe the development of the new AFE software (Chapter 6), with a particular emphasis on the operation of the TL loop (Chapter 7). Gain variations as a function of temperature and software parameters were evaluated using the Compton Edge (Chapters 4 and 7). All measurements were conducted in a climate chamber providing full control over ambient temperature, using a measurement Equivalent Detector (ED), imitating the oryginal MCORD detector (Chapter 2), consisting of small scintillator tiles, AFE electronics, SiPM sensors, and NA-22 particle sources. Finally, in Chapter 8, we present a comparison of system performance with and without the TL loop and assess the effect of different TL parameters on detector stability, identifying those crucial for its optimal operation.

## 2. Measurement system and procedure

The measurement setup used for the planned experiments is shown in Figure 1. Environmental tests with controlled temperature changes were carried out in a Binder MK53 climate chamber [9] with internal dimensions of 40 x 40 x 33 cm³ and a temperature range from −40 °C to +180 °C. Measurements to determine the proportionality coefficient (dV/dT) in the temperature loop (see Chapter 6) were performed in the temperature range between 10°C and 35°C, but measurements to check the operation of the temperature loop were performed at temperatures between 15°C and 30°C. The chamber allowed fully automated execution of pre-programmed measurement sequences, including controlled temperature ramps at specified rates. Temperature changes during each measurement cycle, lasting from several hours to multiple days, were entirely managed by the chamber's internal controller. The computer was connected only to upload the measurement program and did not directly control or monitor the temperature during the experiments.

Due to the large size of a single MCORD detection desk [3], it was not possible to place it directly inside the climate chamber. To overcome this limitation, a special smaller-scale Equivalent Detector (ED) was prepared. Its construction is described below. The ED is housed in an aluminum enclosure, mimicking the original MCORD desk housing, and the entire assembly is placed inside the climate chamber. Measurement and control signals exit the chamber through a single cable connected to the MCORD HUB. The physical signal (analog measurement data) is transmitted without modification — signal amplification is performed inside the chamber by the AFE electronics — before reaching the ADC and analyzer (CAEN DT5730SB [10]), where it is digitized, initially analyzed, and stored. After each measurement, the collected data are transferred to a PC for offline processing and further analysis. A block diagram of the physical signal path is shown in Fig. 2.

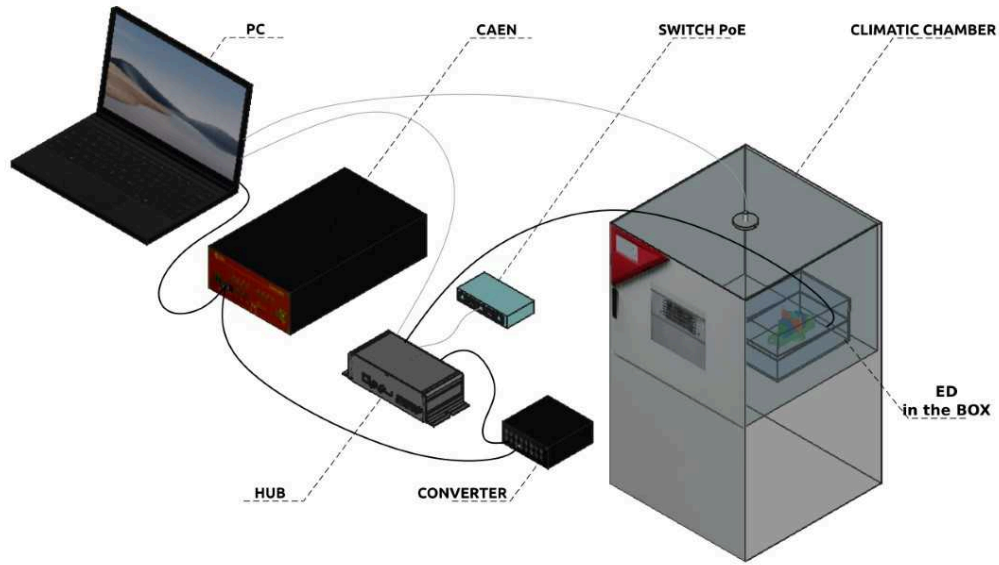

Figure 1. Visualisation of the measuring system. The thick line shows how the measurement signal travels from the analogue front end (AFE) in the Equivalent Detector (ED) to the computer, which records the measurement results. The other lines show that the devices in the system can be configured using a computer or via an Ethernet network.

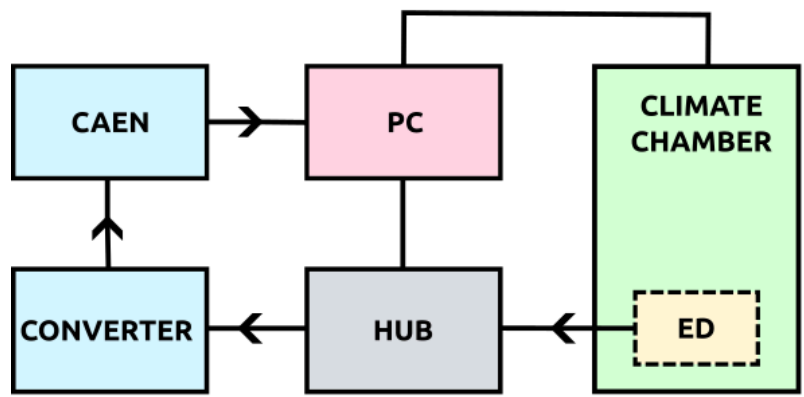

Figure 2. Schematic diagram of the measurement system. The lines with arrows show how the measurement signal travels from the analogue front end (AFE) in the Equivalent Detector (ED) to the computer, which records the measurement results. The other lines show that the devices in the system can be configured using a computer.

As described above, for measurements in the climate chamber, a special small-scale detector model (Equivalent Detector, ED), consisting of two detectors and a radiation source holder (which can hold 1 to 3 sources) located exactly in the center between them, was developed (see 3D cross-sections in Fig. 3). The ED was 3D-printed, allowing full adaptation to the dimensions of the mounted components. It accommodates two small electronic boards equipped with temperature sensors and SiPMs at opposite ends (shown in dark green), small cylindrical scintillators made of the same material as used in the MCORD detector (light blue), and gamma radiation sources positioned along the central axis (red). For these tests, gamma radiation sources were used instead of cosmic rays due to the extended measurement times and the small scintillator dimensions. Na-22 sources with an activity of 1 MBq (in 2018) were employed. These sources emit pairs of particles in opposite directions, ensuring that both scintillators receive simultaneous interactions, and both SiPM sensors (master and slave) register coincident light signals. The ED design allowed for precise positioning of the radioactive sources along the measurement axis and, if required, adjustment of the distance between the scintillators using screws (shown in grey). The ED assembly, including the SiPMs and two AFE electronics boards (master and slave), was enclosed in an aluminum housing, with internal cabling and an external signal cable identical to those used in the original MCORD detector.

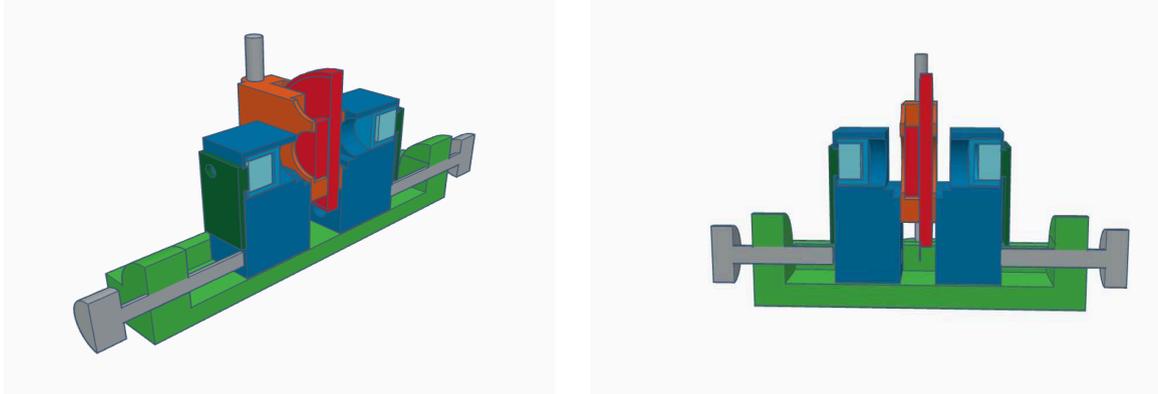

Figure 3. A three-dimensional ED model. The following symbols are used in the figure: SiPM sensors placed on electronic boards - dark green planes, scintillators - light blue cylinders, Na-22 radiation sources - red discs, mounting - light green; clamping screws - gray, mountings for radiation sources - orange, detector mounts - dark blue.

SiPM sensors are highly sensitive to changes in temperature, which directly affect the gain of the sensor signal. To filter out noise, we have clipped the signal at the lower end at 300 ADC channels and at the upper end at 900 ADC channels. Consequently, the proposed detector must maintain a stable SiPM gain regardless of ambient temperature fluctuations. To evaluate both the gain variations of the SiPM photodetectors and the behavior of the scintillators under changing temperatures, the Compton edge in the energy spectrum recorded from the scintillators was used throughout the detector (see Chapter 4). Based on this approach, a dedicated Temperature Loop (TL) was implemented in the detector control software. The TL loop adjusts the SiPM bias voltage to ensure that the position of the Compton edge remains stable, independent of temperature changes. The TL software continuously measures the temperature near each SiPM photodetector, averages the readings, and, when the temperature change exceeds a predefined threshold, corrects the sensor's bias voltage by the required amount. The magnitude of this correction depends on a calibration coefficient, which is determined as described in Chapter 3. A detailed description of the low-level software implementation and the operation of the TL loop can be found in Chapters 6 and 7.

## 3. Temperature Coefficient factor measurement

One of the key parameters of silicon photomultipliers (SiPMs) is their temperature coefficient, which defines the change in breakdown voltage as a function of temperature. Even small temperature variations can lead to noticeable changes in the breakdown voltage and, consequently — if the bias voltage remains constant — to variations in the SiPM gain. In the MCORD detector, Hamamatsu

S13360-3075PE photodetectors (3 mm × 3 mm, pixel size 75 μm) [6] were used. According to the manufacturer, the nominal temperature coefficient is 52 mV/°C. To verify this value for our sensors, a series of measurements was carried out for two samples from the delivered batch (SN4680 and SN4683). The current–voltage (I–V) characteristics were measured using a Keysight B2900A source-measure unit [11], with the sensors placed inside a Binder MK 53 climatic chamber. The chamber temperature was controlled with an uncertainty of ±0.2°C. For sample SN4680, the measurements were performed in the temperature range from 0°C to 30°C in 2.5°C steps, while for sample SN4683, the range was from 0°C to 25°C in 5°C steps. The obtained I–V characteristics are presented in Figure 4. In the figure, the x-axis corresponds to the voltage applied to the SiPM, while the y-axis represents the square root of the dark current measured at a given bias voltage. As shown, the dark current increases sharply once the bias voltage exceeds a certain value, corresponding to the onset of avalanche multiplication. The data points above this voltage (marked in blue) can be well fitted with a straight line. The breakdown voltage $V_{br}$ was defined as the intersection point of this line (red line) with the voltage axis. The dependence of the breakdown voltage on temperature is shown in Figure 5. From a linear fit to the data, the measured temperature coefficient was determined to be 48 mV/°C. The errors of the results presented here were determined by calculating the standard deviation of the series of measurements performed in the same conditions.

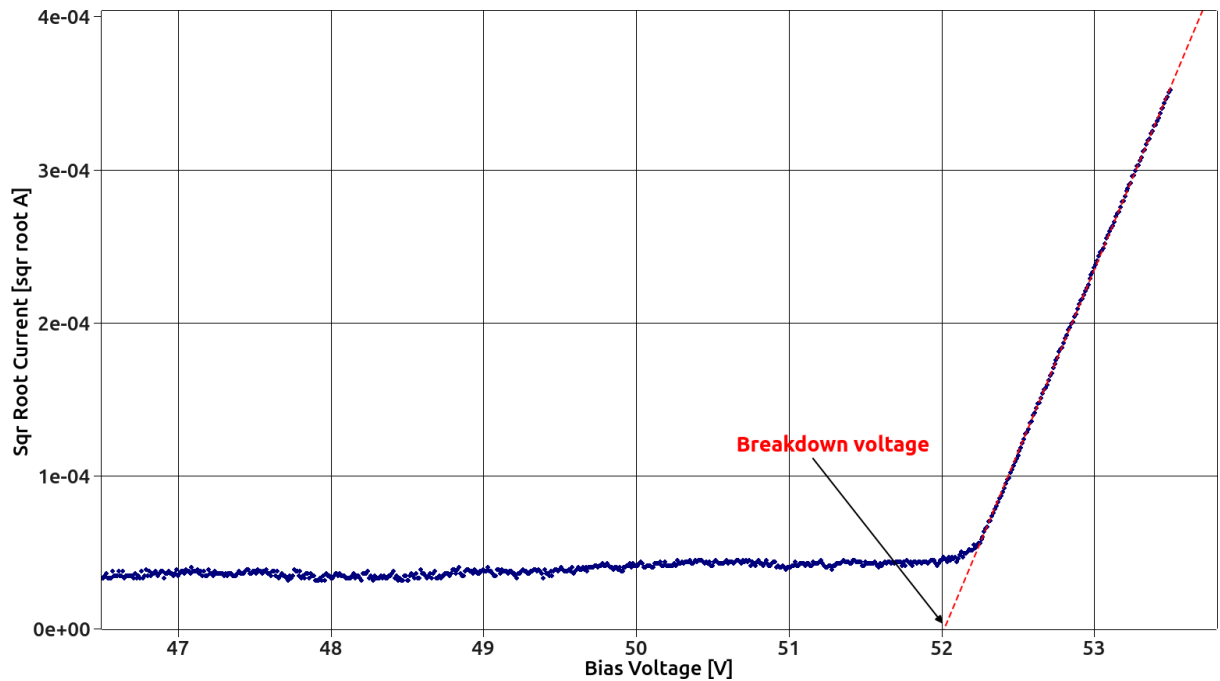

Figure 4. Current–voltage (I–V) characteristics

As can be seen from the laboratory measurements, the supply voltage correction factor we obtained is only slightly lower than the manufacturer's specification (52 mV). Furthermore, since correcting the supply voltage by 1-2 mV is practically impossible, based on these two values, we can assume that our factor is approximately 50 mV per step. The above measurements and analyses focused solely on the SiPM sensor itself. In reality, it is part of a measurement system consisting of the AFE electronics and a scintillator. For this reason, a series of additional measurements were performed to test the behavior of the entire system, including the SiPM sensor, in the same climatic chamber to receive MCORD full aperture/device temperature sensitivity.

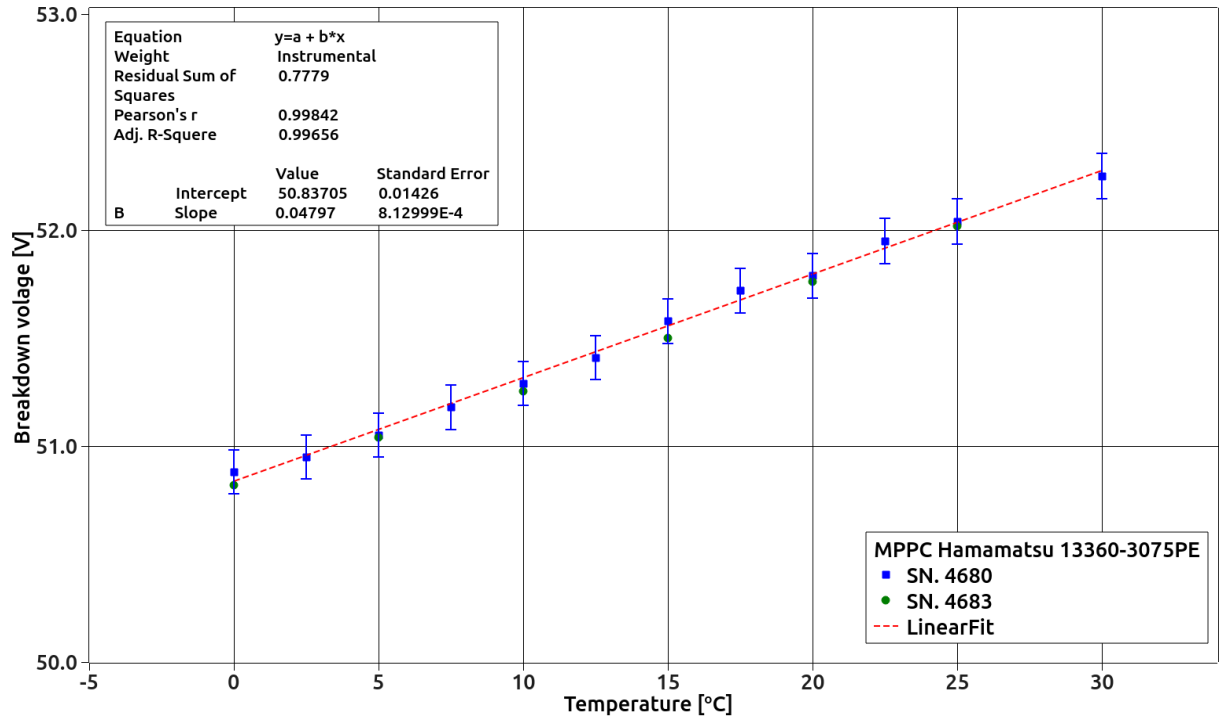

Figure 5. Breakdown voltage vs. temperature.

The sensitivity of the MCORD system changes with temperature variations, not only due to the SiPM sensor itself. In real-world conditions, the presence of the plastic scintillator and the entire AFE electronics must also be taken into account. This set operates as a complete measurement system in close proximity to the SiPM sensor and is affected by the same temperature changes. Testing entire MCORD detector boards under controlled temperature conditions was not possible due to the size limitations of the climatic chamber used. Therefore, a replacement ED set (described above) was developed, the size of which allowed it to be placed inside the climatic chamber. Furthermore, to better match our model to the actual detector, this new set was placed in an aluminum box (replacing the MCORD detector board housing) during chamber measurements.

Measurements of breakdown voltage as a function of temperature were used to determine the correction factor for the SiPM sensor itself. We believe that for the entire system, a good method for determining the correction factor is to measure the coincidence change in the position of the Compton edge as a function of temperature in the spectrum of radioactive Na-22 (annihilation peak from $\beta+$ decay).

Due to the low event statistics (despite using two sources simultaneously), a novel method for determining the Compton edge value and estimating measurement error with low measurement statistics was developed and used (see Chapter 4, Compton Edge). The ED model of the MCORD detector was tested over a temperature range of 15-35°C. Within this range, the value of the SiPM supply voltage change was tested to maintain the reference Compton edge level for a system composed of two sensors (master and slave—the two ends of the MCORD detection board).

The measurement procedure was as follows. Measurements were taken at the climatic chamber temperature points of 10, 15, 20, 25, 30, and 35°C. The system is assumed to operate at room temperature, defined as a climate chamber setpoint of 20°C. Due to the heat generated by the nearby AFE electronics, the temperature measured directly on the SiPM sensor board is approximately 23°C. This temperature is taken as the reference operating point in the subsequent analysis. The differences between these temperatures are due to the fact that the AFE electronics generate heat during operation and, in addition, the system is housed in an aluminum enclosure. This difference after the temperature has stabilized following a change is roughly constant across the entire range of temperatures tested and amounts to approximately 3.5°C for the SiPM Master and 2.5°C for the SiPM Slave (Fig. 6). The graph shows that after a change in the chamber temperature, it took a longer time, approximately 0.5 hours, for the temperature inside the ED enclosure to stabilize. A similar phenomenon occurs in the actual MCORD detector board. In the remainder of this article, we will primarily use the temperature value reported by the AFE sensors.

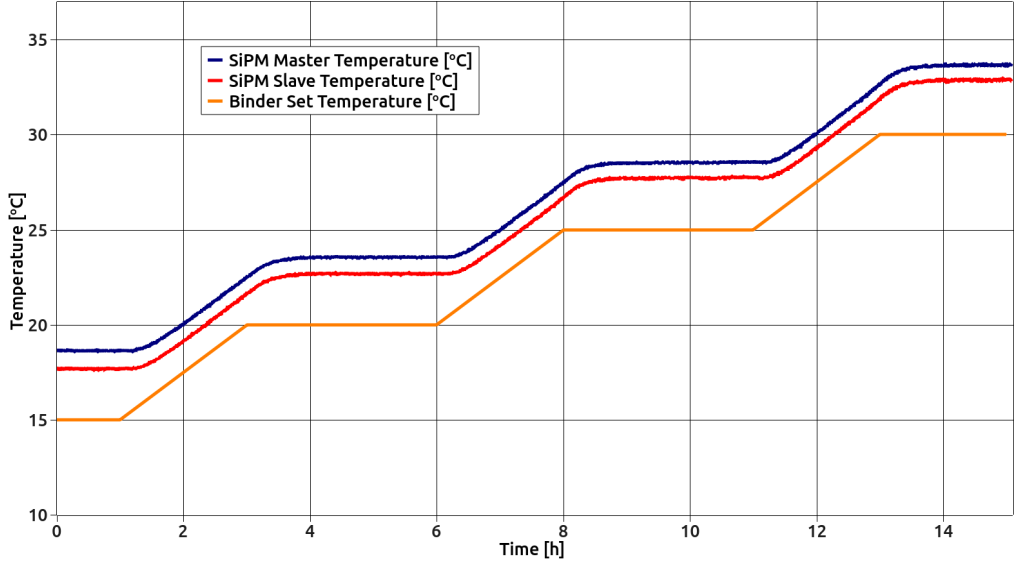

Figure 6. The graph compares the temperature set in the climate chamber and the temperature read by the SiPM sensors.

As the temperature in the climate chamber increased at a constant supply voltage, the Compton edge value decreased (the energy spectrum shifted to the left towards lower values). Figure 7 shows a straight line fitted to the points determining the relationship between the Compton edge value (expressed in ADC Chanel) and the voltage applied to the SiPM at 25°C. The figure also shows the Compton edge value at 20°C at a voltage of 53.5 V for the SiPM Master and 54 V for the SiPM Slave (these values will be referred to as reference values later in the article). By taking several measurement points around the reference value and applying linear regression, the voltage required to obtain the Compton edge reference value was determined with satisfactory accuracy. We consider the Compton edge values determined in this way to be optimal. Therefore, we want to select the voltage at a different temperature (the figure shows an example relationship for 25°C) so that the Compton edge does not change. To this end, we determine the linear relationship between the voltage value on the SiPMs (Master and Slave) and the Compton edge for a given temperature and check for which voltage this edge remains the same as the reference value. The calculated voltage is expressed using the following equations:

$$U_K = (K_{ref} - b)/a, \qquad (1)$$

where $U_K$ is the voltage required to maintain the Compton edge at a constant value, $K_{ref}$ is the reference value of the Compton edge, and *a* and *b* are the coefficients of the linear fit of the Compton edge versus voltage relationship for a given temperature. These voltages are marked in the figure with the symbols $U_{KM}$ for the Master SiPM and $U_{KS}$ for the Slave SiPM. The figure shows the case of the Master SiPM when, after changing the temperature from 20°C to 25°C, in order to maintain the same Compton edge value (768.1 ADC channels), it is necessary to increase the voltage from 53.5 V to 53.9 V ($U_{KM}$). It also shows the case of a Slave SiPM, where when the temperature changes from 20°C to 25°C, in order to maintain the same Compton edge value (716.3 ADC channels), it is necessary to increase the voltage from 54.0 V to 54.3 V ($U_{KS}$).

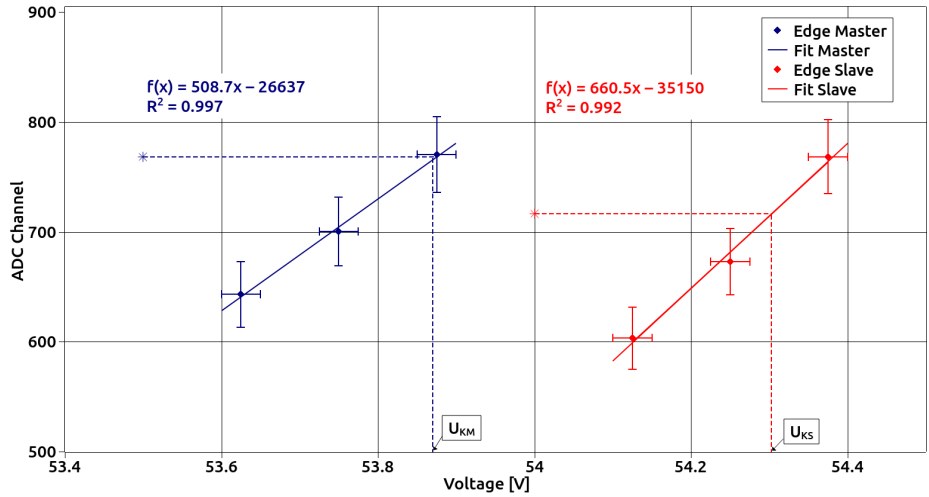

Figure 7. The graph of Compton edge values versus voltage.

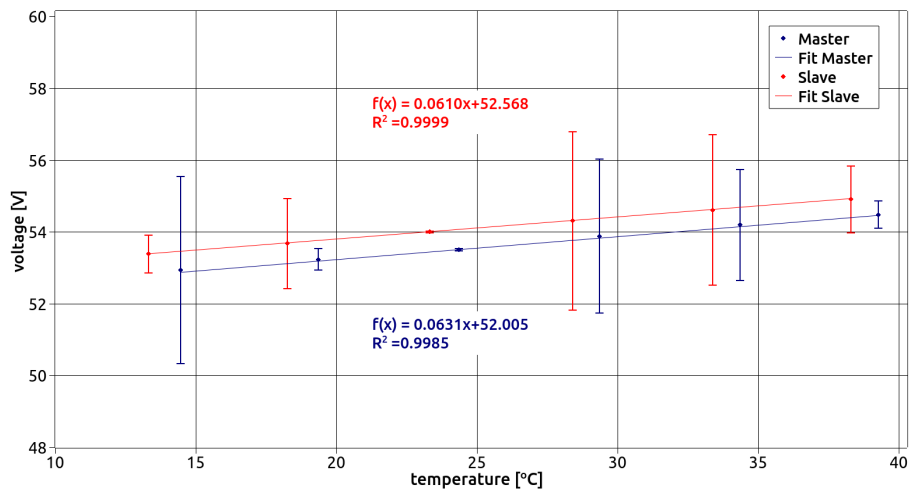

Figure 8. Determination of the temperature coefficient

Voltage uncertainty (see Figure 7) was determined by comparing the AFE circuit with a Keithley 6517A precision electrometer/voltmeter [7] and was found to be 0.025 V over the entire range of measured voltages. The uncertainty with which the Compton edge position was determined is described in detail in Chapter 4.

This series of measurements was performed for six consecutive temperature changes, and the resulting voltage correction values were plotted on a single graph (voltage versus AFE temperature – Fig. 8). In this graph, the slope of the linear function, adjusted for the six measurement points, is the desired coefficient factor for the entire system. Measurements were collected separately for the AFE master and slave electronics, but ultimately, a single average value was adopted. The measurement results shown in Figure 8 show that the coefficient value for the entire system (SiPM, scintillator, AFE electronics, housing) is significantly higher than the value determined for the SiPM alone (as discussed in Chapter 3) and is approximately 62 mV/°C. This coefficient value was used in all further measurements. The measurement uncertainties associated with the temperature measurement are 0.2°C. The measurement uncertainties associated with voltage correction (except for the reference value) were calculated using the total differential method based on the uncertainty of determining the reference Compton edge value and the calculated linear regression fit coefficients to points near the reference voltage (see Figure 7). The formula for the uncertainty of the voltage calculated using this method is as follows:

$$\Delta U_K = \left|\frac{1}{a}\right|\Delta K_{ref} + \left|\frac{K_{ref}}{a^2}\right|\Delta a + \left|\frac{1}{a}\right|\Delta b \qquad (2)$$

where $\Delta U_K$ is the uncertainty of the voltage being sought, $a$ and $b$ are the coefficients of the linear fit of the Compton edge versus voltage relationship for a given temperature, $\Delta a$ and $\Delta b$ are the uncertainties of these coefficients, $K_{ref}$ is the reference value of the Compton edge, $\Delta K_{ref}$ is the uncertainty of this reference value. The measurement uncertainty for the reference voltage arises solely from the accuracy of the voltage measurement by the AFE and has been estimated at 0.025 V. The measurement uncertainty of the temperature coefficient was calculated using a linear regression procedure and has been estimated at 1 mV/°C.

## 4. Compton edge and error analysis

To determine the Compton edge by comparing measurement results with various analytical functions, we assumed that the following function (see eq. 3) fits the measurement results well enough to determine the parameters for the voltage correction procedure for the SiPM sensors in our ionizing radiation detector:

$$I(E) = \begin{cases} \frac{a-c}{b^2}E^2 - 2\frac{a-c}{b}E + a & \text{dla} \quad E \leq b \\ c & \text{dla} \quad b < E \leq d \\ -\frac{c}{e-d}E + \frac{ce}{e-d} & \text{dla} \quad d < E \leq e \\ 0 & \text{dla} \quad E > e \end{cases} \qquad (3)$$

where the function parameters were chosen so that on the boundaries of the first (E <= b ) and second (b < E <= d ) intervals, both the function values I(E) and their derivatives I'(E) were equal. For the remaining interval boundaries, we assumed that the condition that the function values were equal on the boundaries of the intervals was met. Next, we assumed that the function interval in which it takes on a constant value c corresponds to a local maximum, and the energy value for the Compton edge lies on the line describing the fitted function for the interval (d < E <= e) and corresponds to a radiation intensity of 75% of the value of the local maximum of the radiation intensity. The parameters of the fitted function are shown in the graph (see fig. 9) below. The next figure (Fig. 10) shows an example of such a fit to real data.

These assumptions mean that the energy of the Compton edge is given by the formula:

$$E_{com} = \frac{3}{4}d + \frac{1}{4}e \qquad (4)$$

The procedure for determining the fitting parameters is implemented using the NonlinearModelFit function in Wolfram Mathematica [12]. It can also be used to obtain the measurement uncertainties of the individual fitting parameters. Knowing these uncertainties, the uncertainty of determining the Compton edge can be calculated using the total differential method.

Applying this method leads to the following formula:

$$\Delta E_{com} = \frac{3}{4}\Delta d + \frac{1}{4}\Delta e \qquad (5)$$

where $\Delta d$ and $\Delta e$ are the uncertainties obtained using the previously mentioned Mathematica function. In the rest of this article, this uncertainty $\Delta E_{com}$ will be referred to as statistical uncertainty.

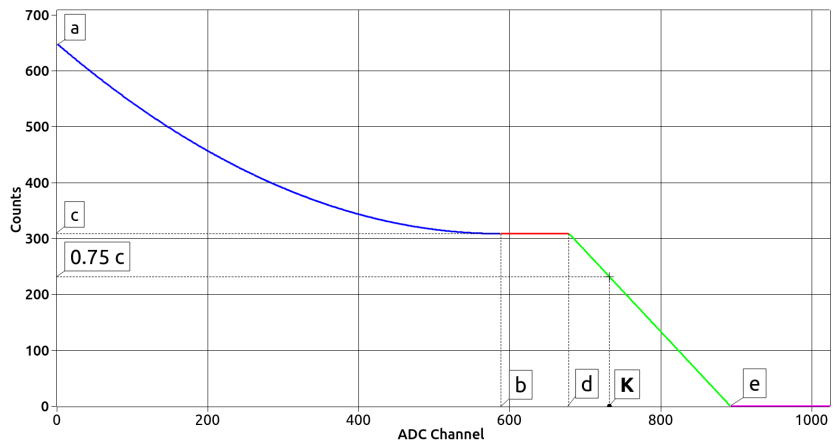

Figure 9. A function defined by Eq. 3 was fitted to the measured data. It consists of four parts - the first part (dark blue line) - defined on the interval (0, b] is described by a quadratic function that starts at the point (0, a) and ends at the point (b, c), then in the interval (b, d] there is a plateau with value c (red line), which on the interval (d, e] turns into a decreasing linear function (green line). We assume that from a certain moment (for values greater than e) the number of detected gamma quanta decreases to zero (magenta line). We assume that the Compton edge energy corresponds to 75% of the plateau with value c and is recorded on the linearly decreasing (d, e] interval of the fitted function - this means that the Compton energy calculated in this way is expressed by Eq. 4 through the parameters of the function (see Eq. 3).

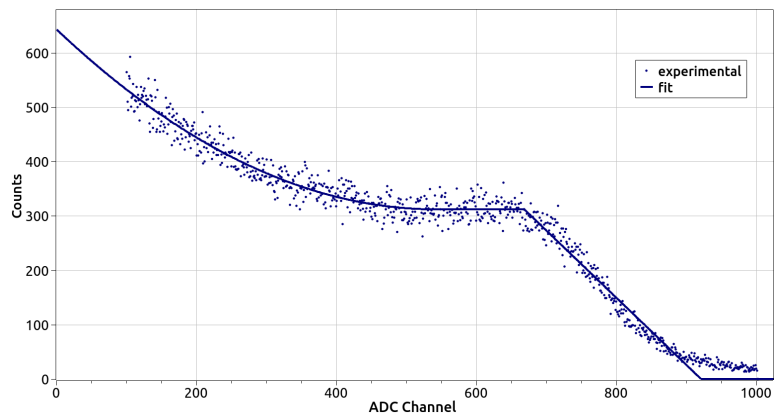

Figure 10. Comparison of measurement data (dots) with the relationship fitted to them (line). The graph shows the typical course of the measured Na-22 coincidence spectrum (511 keV line) for our detector and the function fitted to it, defined by the formula Eq. 3 in case of very long (2 days) measurement

In addition to the aforementioned statistical uncertainties resulting from the Compton edge determination method, there are also uncertainties related to the limited precision of the voltage setting on the SiPM and the limited accuracy of the temperature reading. The errors in determining the temperature and voltage are small compared to the other errors and they are included in the complete error described at the end of this chapter.

4.1    Dependence of the statistical error limit on the measurement time

The measurements were made using Na-22 sources that were several years old. Their effective intensity was significantly lower than the original, so we used two or three sources simultaneously to improve statistics and shorten the measurement time. To determine the recommended minimum measurement time, a series of identical measurements were performed, varying only the duration.

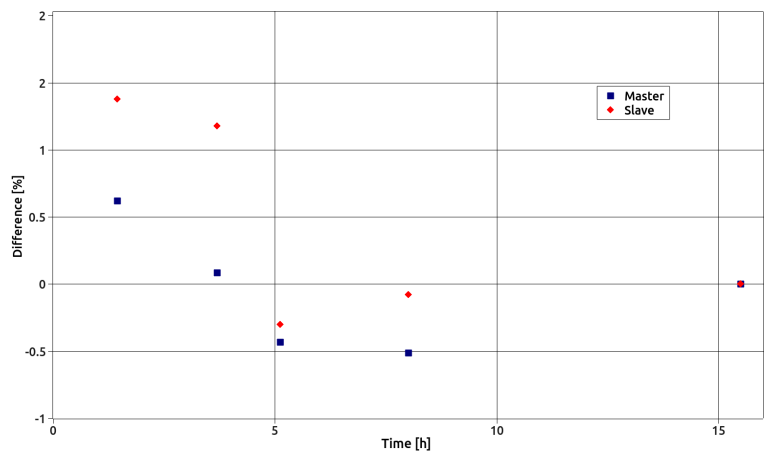

Figure 11. Measurement at a constant temperature of 25°C. The graph shows how the Compton edge value converges over time to a value determined after approximately 16 hours (the y-axis shows the percentage difference between the value measured after 16 hours and the values measured in earlier time intervals).

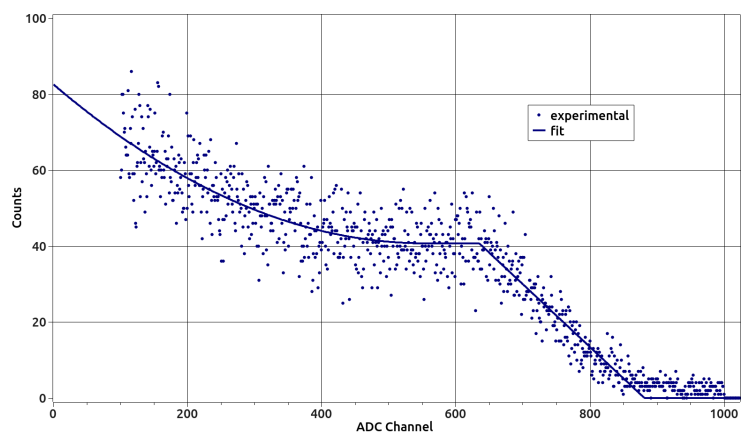

Figure 12.  The fitting method is performed in case of a short (8 hours) measurement.

Figure 11 shows the results of these measurements after 1, 4, 5, 8, and 16 hours, with the 16-hour measurement taken as the reference level. We see that after 6-8 hours, the error rate drops to less than 0.5% compared to much longer measurements. In subsequent measurements, we assumed this as the minimum measurement time.

It should be noted, however, that after 8 hours of measurement, the number of counts near the upper plato of the fit was only about 30-40. Our fitting method performed very well even for such low measurement statistics, as seen in Figure 12. However, to increase measurement confidence, we ensure that most measurements have significantly better statistics than 100 on the upper plato (see Figure 10).

4.2 Repeatability of measurements

To verify the repeatability of the results in our measurement setup, 11 identical measurements were performed. Each measurement lasted 6 hours (corresponding to the collected statistics at the upper plateau of greater than 30 counts) and was performed at a constant temperature of 30°C. As can be seen in Figure 13, the dispersion of the measurement points is relatively small, resulting in an error of approximately 0.5%. For the AFE Master electronics, the dispersion was 0.42%, and for the Slave electronics, 0.55%. We can see that the repeatability of the results is good even with low statistics.

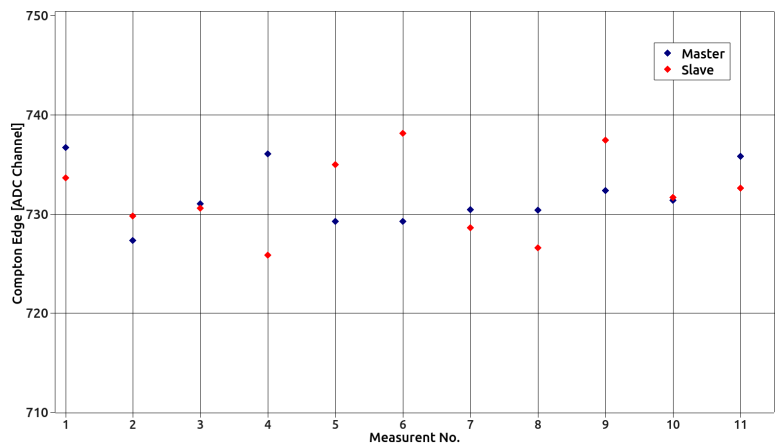

Figure 13. Measurement of the Compton edge at a constant temperature of 30°C in a climate chamber. The same six-hour measurement was repeated 11 times. It can be seen that the differences between the measurements are small and that there is a high degree of repeatability between measurements.

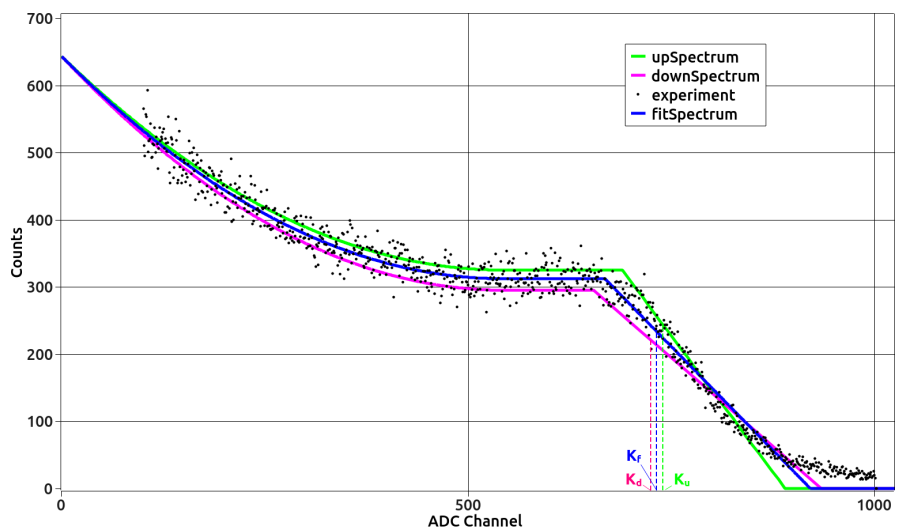

Figure 14. Three types of numerically fitted curves to measurement data (using Mathematica). The difference between $K_u$ and $K_d$ is the measurement uncertainty for "Manual Parameter Adjustment" estimated by us.

4.3   Measurement uncertainty arising from differences between repeated measurements carried out under similar conditions

We estimate this uncertainty based on the average differences for a different series of measurements. Some of the measurements, which initially lasted several hours, for better statistics were repeated with a measurement time of 72 hours. These measurements were taken after removing the source from the climate chamber and ED, then repositioning it in the ED and placing it back in the climate chamber. We believe that if we take the average of the differences between the Compton edges determined in this way, it will be a good estimate of the impact that minor imperfections in sample positioning and other factors may have. The uncertainty of Compton Edge $\triangle$E value determined in this way is approximately 12 ADC channels (1,91 - 2,29%).

4.4   Measurement Uncertainty Based on Manual Parameter Adjustment.

Another contribution to estimate measurement uncertainties is to manually determine the extreme parameters for one of the measured curves, for which the fit to the measurement data seems reasonable. Figure 14 shows a comparison of the curve described by Equation 1, numerically fitted to the experimental data (see Chapter 4), and two functions manually fitted in such a way that they take extreme but reasonable-looking values below and above the numerically fitted function. Based on these parameters, two Compton edge values ($K_u$ and $K_d$) were calculated, and the difference between them was used as the basis for calculating the relative uncertainty. This value, expressed as a percentage (2.27%), forms the basis for calculating the measurement uncertainty for all measured relationships.

4.4   Summary of measurement errors

We assumed that the uncertainty with which we measure the Compton edge is the sum of the difference between measurement results repeated under very similar conditions and the statistical uncertainty arising from the fact that the fitted function parameters (Eq. 1) are subject to uncertainties, as well as a manually determined uncertainty based on an assessment of how reasonable the fit appears for the various parameters.

## 5. MCORD electronic correction

As shown in article [7], the AFE electronics we used so far and the converters ADC within it generated a certain level of noise (up to 20-30 bits), which prevented precise determination of the supply voltage. We decided to modify the existing electronic circuit to mitigate this effect. The changes described in this chapter are supplementary to the information provided in article [4], and we will not refer to this topic in the rest of this article.

During calibration measurements of AFE circuits for very low SiPM current levels (on the order of $1.0 \cdot 10^{-7}$ A, corresponding to values of several tens of bits on the ADC), significant signal fluctuations were observed, characterized by a standard deviation at the level of several bits on the ADC. This resulted in poor repeatability of results at very low currents. After analyzing the AFE electronic circuit, we concluded that the cause was too short averaging time of the acquired signal at the U3 operational amplifier (Fig. 15), which is part of the LDO block (Low DropOut regulator - built-in linear voltage regulator with low voltage drop). To improve measurement accuracy, capacitors C88 (1 µF) and C152 (10 µF) were added to the LTC6101HV operational amplifier circuit. This modification resulted in an increased measurement time constant and a substantial reduction of current fluctuations. The standard deviation decreased to below 1 bit.

For low current levels, the standard deviation was comparable to the measured signal value, whereas after adding the capacitors, it decreased to a few percent of the measured value. In Figure 16 Left, we see that the mid- and high-range current measurements were equally accurate before and after the changes (the red and blue points overlap and form lines). However, for low currents, the spread of the blue points (the system before the change) is much greater than that of the orange points (the measurement after the change). Additionally, for low current intensities, which are particularly important when determining the breakdown voltage of SiPMs, the introduction of the capacitors significantly enhanced the repeatability of calibration measurements (Fig. 16 Right).

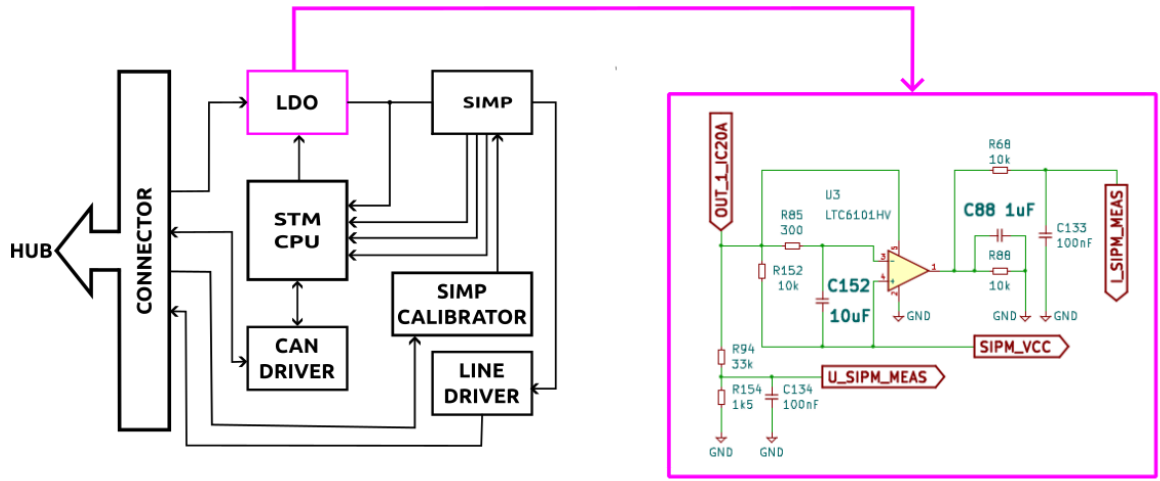

Figure 15. Block diagram of the AFE circuit with modifications marked (bold text) in the electronic diagram of the LDO block.

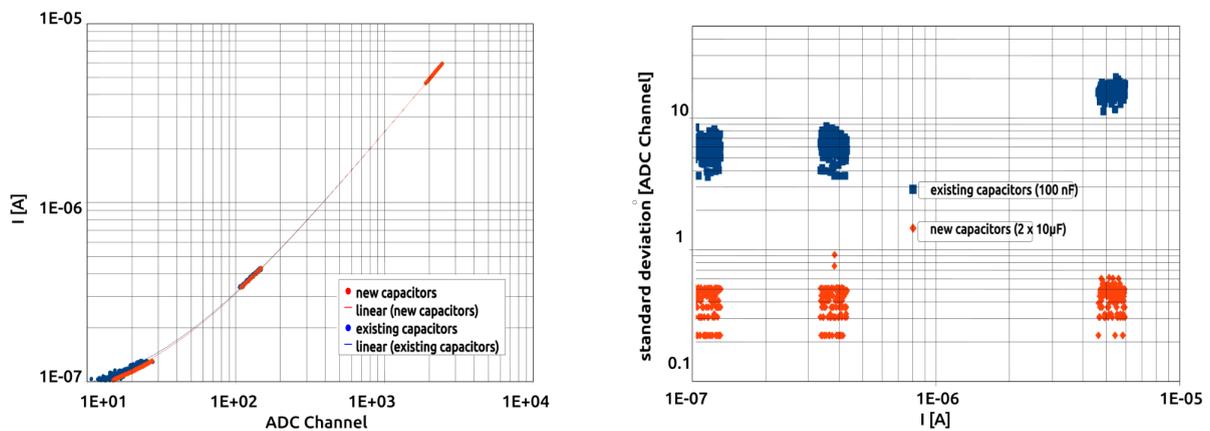

Figure 16. Left: The relationship, measured for the purpose of calibrating the AFE, between the current measured by the AFE (the average of several measurements in the ADC channels) and the current calculated using voltage measurements on three reference resistors – a detailed description can be found in the article [7]. Right: The standard deviation from the mean current value for measurements at various current levels.

As a result of using an additional capacitor in our circuit, the noise level decreased by a factor of 10 and current measurements performed using an internal meter are much more accurate and, most importantly, the measurement values for small currents are repeatable.

# 6. Update of AFE and HUB software

To date, measurements using the MCORD detector have utilized electronics software written by individuals from the Warsaw University of Technology during the project's initial stages (see article [7]). After consultations, it was decided to not only modify the temperature loop portion of the software but also rewrite the entire software, including the detector activation section.

6.1 HUB Firmware and control

The firmware is written in micropython [13]. Each task is run by the UASYNCIO library, to achieve multitasking. By default, *"Read, Eval, Print, Loop"* (REPL) for optimization, but it can be enabled. HUB has an SD card, where loaded firmware and data from AFEs can be stored. HUB can be controlled via ethernet, thanks to the server task. If AFE is found, then configuration is read from the configuration files. HUB has turned on Independent WatchDoG (IWDG).

The HUB Main Loop is dedicated to manage AFEs. It is divided by subtasks:

- dequeue CAN message – it gathers the last message (if available) from the CAN Handler task.
- discover devices – it trying find all connected AFEs (by sending 0x00 command)
- process received message – it processing received messages, obtained by first subtask
- managing AFE – it manages each discovered AFE

The HUB server manages communication over ethernet. The request, for early development, is in JSON format. It should contain the name of the procedure (e.g. close all, get all configuration, power off, set parameter, set voltage) and its values. After receiving such a request, the client should wait for the response. Sometimes it can take a long time (a few seconds) due to the other tasks and on HUB and on AFE. The periodic task contains functions to manage logger (saving data on a SD Card) and to print by UART (USB).

6.2 AFE Firmware overview

The AFE Firmware is written in C with the STM32F0 HAL library [14]. It operates as a state machine that continuously monitors for incoming CAN messages and measurements. All interaction with the AFE module is performed via the CAN bus. The device uses standard CAN 2.0A frames with 11-bit identifiers. For each of the 8 ADC channels, the user can independently configure data processing parameters. Upon receiving a valid command, it performs the requested action – such as reading last measurement, setting DAC or updating configuration parameters. The firmware also manages several autonomous tasks, including:

- Periodically acquiring data from all configured ADC channels and storing it in circular buffers.
- Executing the temperature compensation feedback loop (Temperature Loop).
- Managing the gradual ramping of DAC Voltages.
- Transmitting periodic status and data messages if configured to do so.
- Listening for incoming CAN messages and executing commands.
- Managing software watchdogs.

The ADC can be triggered either by a configurable TIM (hardware timer) with DMA (Direct Memory Access) or by software. The previous version was triggered by software (polling), and calibration was performed for that option. After a software update, calibration differed for the ADC when using DMA, so the ADC is now triggered by software (polling).

The new polling method allows other tasks to continue during polling, but it can be enabled if communication is more critical than measurements. During each ADC polling cycle, every ADC channel is measured. However, the timestamp and raw measurement value are stored in the circular buffer only at the programmed intervals.

6.3 Temperature Loop Software

A critical feature for SiPM operation is the automated temperature compensation loop. The gain of a SiPM is linearly dependent on the over-voltage, which in turn depends on the breakdown voltage. Since the breakdown voltage varies with temperature, the bias voltage must be adjusted to maintain a constant gain. The AFE module implements a closed-loop feedback system to perform this compensation automatically. The user configures the loop with the following parameters, independently for each channel:

- **Temperature Coefficient (dV/dT)**: The change in bias voltage required per degree Celsius change in temperature (V/°C).
- **Optimal Operating Point (T_opt, V_opt)**: The desired reference temperature and the corresponding optimal bias voltage.
- **Dead-band (ΔT)**: A temperature window around the last set point within which no voltage adjustment is made, preventing oscillations.

The firmware periodically reads the temperature from a designated sensor channel, compares it to the last recorded temperature, and if the change exceeds ΔT, it calculates the new required bias voltage using the formula:

$$V(T) = V_{opt} + \frac{dV}{dT}(T - T_{opt}) + V_{cor}, \qquad (6)$$

where $T$ is the averaged temperature and $V_{cor}$ (manually added offset voltage value - default value is zero). The DAC output is then updated to this new value. The entire loop can be enabled or disabled.

To protect the connected detector from sudden voltage changes, the firmware includes a DAC ramp controller. Instead of setting the DAC output directly, the user can set a target voltage. The firmware will then automatically ramp the voltage from its current value to the target value in small, configurable steps. The ramp rate (e.g., step size and time between steps) can be configured, ensuring a smooth and safe transition. This feature is active for both manual voltage changes and adjustments made by the automated temperature compensation loop. The default value of the ramp is set to 100 bits per 100 ms, but it can be changed by CAN commands.

For monitoring purposes, any ADC channel can be configured to periodically transmit its latest or averaged reading over the CAN bus without requiring a request from the host. The user can set the reporting period for each channel individually. This reduces the CAN bus load in systems where continuous polling is otherwise necessary.

## 7. Temperature loop

While establishing the operating assumptions for the entire AFE software (described earlier in Chapter 6), a definition of the temperature loop operation was also prepared, i.e. how the supply voltage of the SiPM sensors will change with the temperature change. We want the voltage setting on the SiPM sensors to be corrected so that the data is on a straight line (Fig. 13):

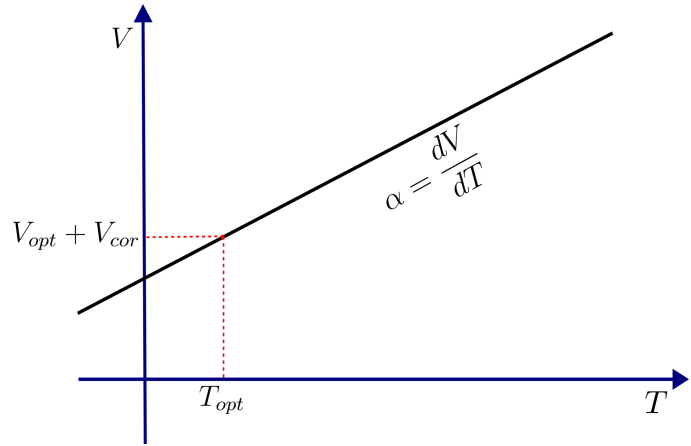

Figure 13. A straight line showing the relationship between the voltage across the SiPMs and temperature, parameterised by the coefficients used in the temperature control loop.

$$V(T) = \alpha T + b, \qquad (7)$$

$$\alpha = \frac{dV}{dT}, \qquad (8)$$

where α is the slope of the line. This coefficient is known as the temperature coefficient and its value can be taken from the manufacturer's technical documentation or determined by independent laboratory measurements (see Chapter 3). The b coefficient is calculated based on the condition that the line must pass through the optimal point:

$$P(T_{opt}, V_{opt} + V_{cor}), \qquad (9)$$

which leads to the equation (6) from chapter 6.3, where dV is the voltage value by which the change should occur after a one-degree temperature change. T is the average ambient temperature of a given SiPM sensor and is determined using measurements taken by temperature sensors located near each SiPM sensor. $V_{opt}$ is the operating voltage determined at temperature $T_{opt}$. The $V_{opt}$ voltage is called the optimal voltage, and the method for determining it was described in article [7]. Currently, the important information for us is that the value of this voltage at a given temperature is given, and our program retrieves it from the database as the reference voltage. It was assumed that the reference/optimal voltage was determined for a temperature of 25°C. If these values for a given SIPM sensor are missing from the database, the program retrieves the average value for the remaining measured values. $V_{cor}$ is the voltage (additional factor) by which we can correct the determined operating voltages if this is necessary for a given measurement (e.g., when the signal is saturated), or in case a systematic error not previously considered is detected.

There are various ways to calculate the average temperature value. Our software will allow for the ability to modify the average calculation method (see section 7.1) to compare the effectiveness of the solutions used. However, our procedure will also consider other parameters related to the average calculation method, such as the number of measurement points from which the average is to be calculated, the time over which these points will be collected, and the temperature change threshold (ΔT) that must be exceeded before the program (temperature loop) performs a voltage correction.

For the temperature loop few averaging methods were prepared: arithmetic mean, weighted exponential moving average, geometric mean, harmonic mean and root mean square. The libraries used in AFE were used to perform the simulation. The simulator incremented the time and used the given function to determine the ADC value:

$$10 \left(\sin{\frac{2t}{T}} + \sin{\frac{t}{T}} + 2 * 0.2 \left(2 U(0,1) - \frac{1}{2}\right)\right) + 100, \quad (10)$$

For T=10000 and given time t in ms. Every step was recorded to the file as an "raw" (ADC sampling) and internally processed data by selected setting for the same time and ADC value. The simulator was intentionally prepared as a technical tool to test the algorithms used in the AFE.

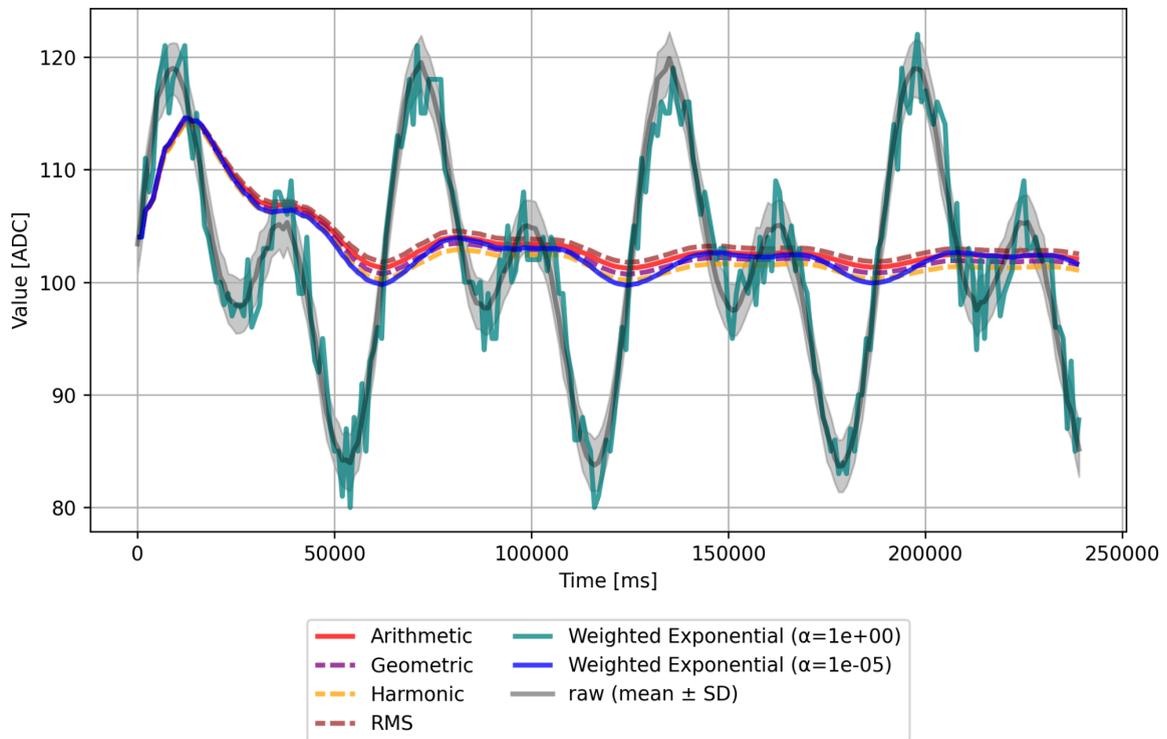

Figure 17. Compare temperature averaging algorithms. Settings: Buffer size: 256, save measurement every 1000 ms, sampling every 10 ms

Figure 17 presents the algorithms used in simulation can be divided by 3 basic groups: arithmetic mean, weighted exponential low and high alpha. In weighted exponential, alpha can modify the importance of the latest measurements, while in other smoothing depends mainly by buffer size and time between measurements. Figure 18 shows the influence of other parameters: buffer size and measurement interval for arithmetic and weighted exponential mean.

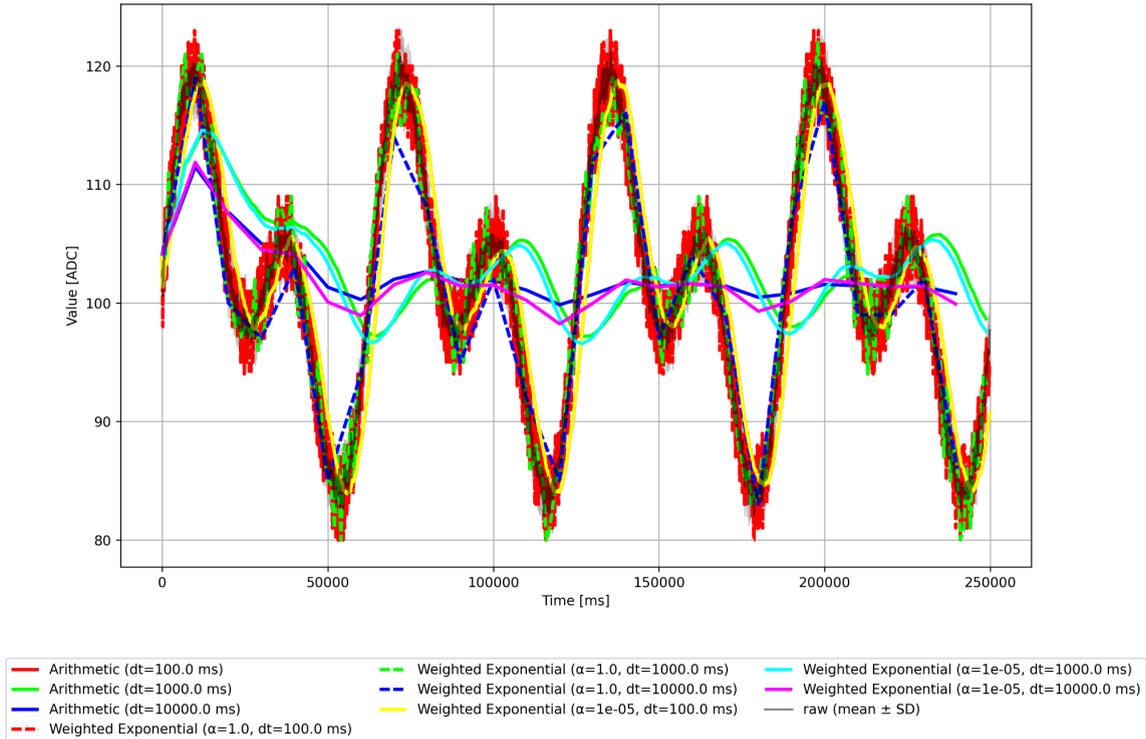

Figure 18. Influence of the parameters on weighted exponential average. Settings: Buffer size: 256, save measurement between 100 and 10000 ms, sampling every 10 ms.

## 8. Measurement results and TL loop analysis

Below, we present the most important measurement results collected during a measurement session lasting several months. This long time was primarily due to the fact that we attempted to perform most measurements with high accuracy (see Chapter 4), which translated into long three-day measurements. By performing subsequent measurements, we tried to verify the temperature loop's performance even with very long measurements, but above all, we tried to assess how various parameters of the loop itself affect its effectiveness. The new software for our electronics (see Chapters 6 and 7) allowed us to adjust a number of variable parameters affecting the temperature loop's performance. We wanted to demonstrate which of these parameters have a significant impact and which are less important. The parameters we adjusted included: the method for calculating the average temperature, the duration of temperature measurements used to calculate the average, and the activation threshold (the minimum temperature change after which the loop corrected the supply voltage of the SiPM sensor). Our software also allows for the adjustment of the number of temperature measurement samples at a given time, but we did not test this variable in these measurements.

After completing preliminary measurements and analyses (see Chapters 3 and 4), the temperature loop's operation was first verified over a multi-day measurement cycle, during which the ambient temperature (inside the climatic chamber) changed repeatedly between 15 and 30°C. Each 5°C temperature change was followed by a period of several hours of constant temperature. Figure 19 shows the time-correlated temperature changes and changes in the SiPM supply voltage. It is also clear that voltage changes occurred only when the cumulative temperature change exceeded a preset threshold (in this case, 0.5°C). This measurement demonstrated that the temperature loop's operation was always correct and stable over time.

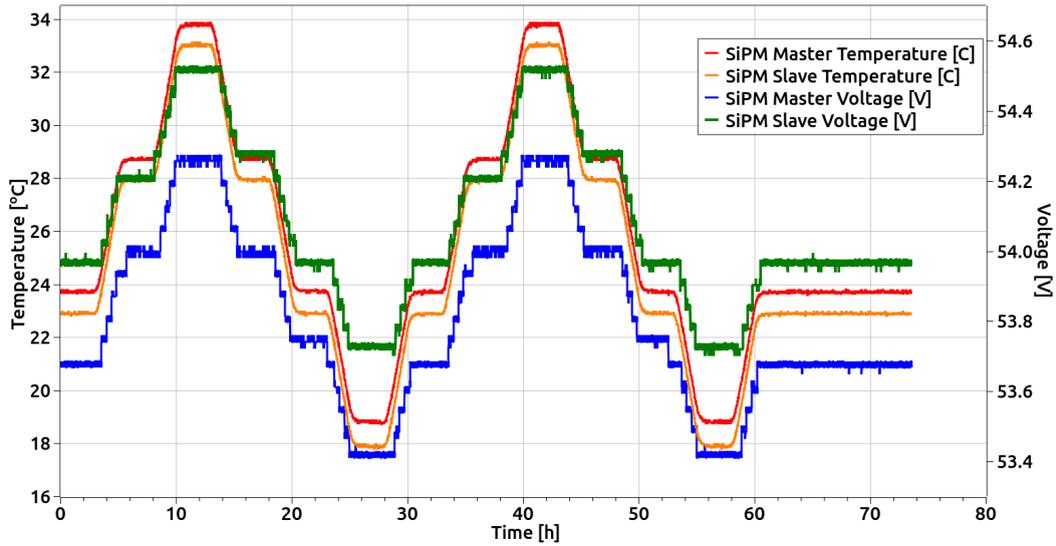

Figure 19. Example of the operation of the temperature loop during multiple temperature changes and stabilization (change of the SiPM sensor supply voltage compensating changes in the ambient temperature), during an example 3-day measurement in a climatic chamber.

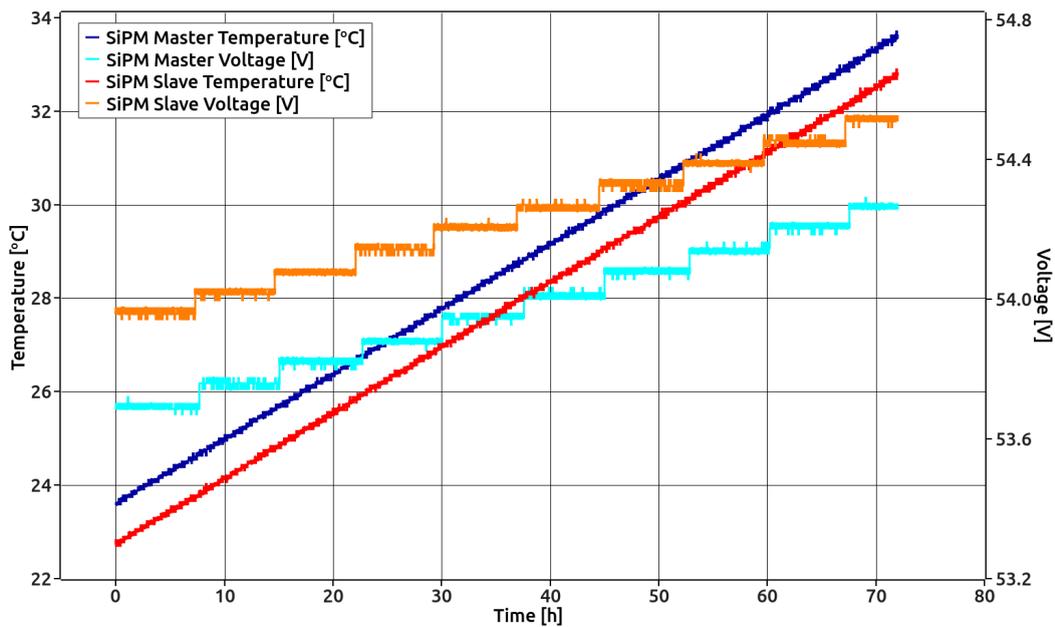

Figure 20. Example of the temperature loop operation during our real measurements (temperature change occurred only in one direction), during a 3-day measurement in a climatic chamber.

In subsequent measurements, the temperature change cycle shown in Figure 19 was not used, but a simpler cycle of only increasing temperature within the 20-30°C range (Figure 20). When we used both increasing and decreasing temperature in a single measurement, the impact of these changes was much more difficult or impossible to observe, as their effects were of opposite sign. The temperature change was uniform and extended throughout the three-day measurement period. Figure 19 shows the effect of the temperature loop's threshold operation (step-wise voltage changes) much more clearly.

To assess the effectiveness of the temperature loop in our cosmic ray detector, we compare the Compton edge value. To facilitate comparison, loop performance measurements were taken at constant temperatures of 20 and 25°C (during this measurement, the loop does not change the supply voltage). Figure 21 shows a comparison of this measurement at a constant temperature of 20°C and two other measurements where the temperature varied according to the cycle shown in Figure 20, with the temperature loop enabled and disabled. The graph shows the qualitative difference between the measurement with the temperature loop disabled and the measurements with the loop enabled. Measurements with the loop enabled at a constant temperature of 20 °C and with the loop enabled at a temperature varying according to program 16 are identical within the margin of error. We can clearly see that the absence of a temperature loop or its malfunction will be clearly visible in the spectrum shape near the Compton edge. Determining the Compton edge value in such a situation is meaningless, and the resulting values will significantly differ from those in a properly functioning detector. From this figure, the measurements no longer show separate curves for the SiPM master and slave sensors (two ends of the detector board), only one example curve.

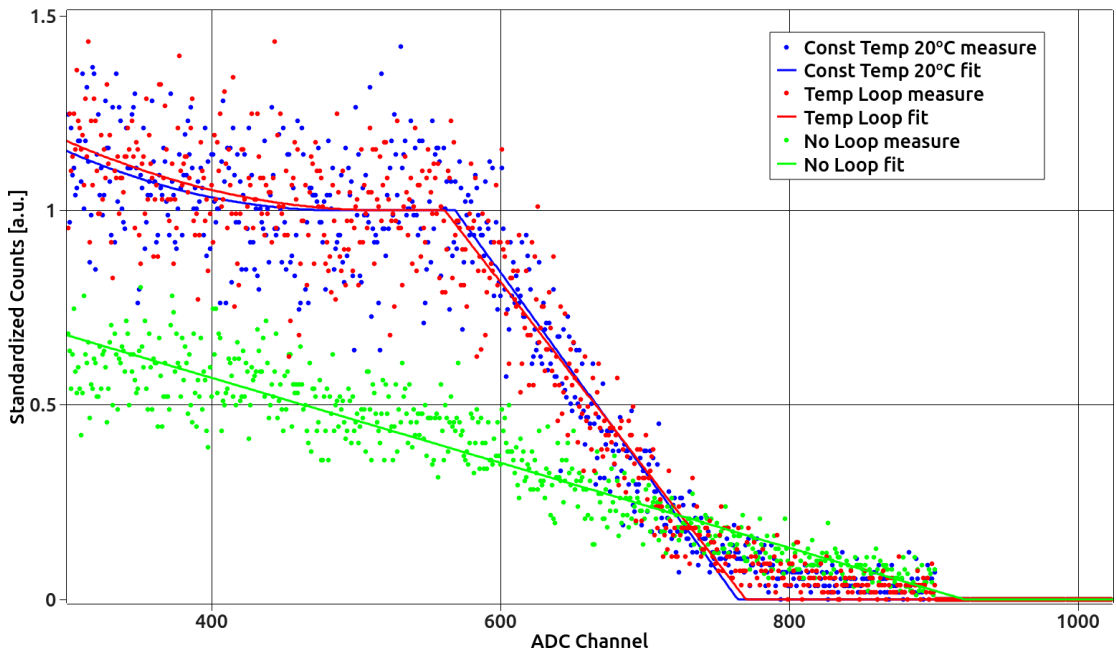

Figure 21. Comparison of the three measurements: with the temperature loop disabled during temperature changes (green); with the temperature loop enabled at a constant temperature of 20 °C (red) and with the temperature loop enabled during temperature changes (dark blue).

Figure 22 shows a series of measurements taken at different threshold temperature values (dead band). The threshold value was varied from 0.5°C to 3°C. For better comparison, the figure also includes a graph of the measurement at a constant temperature of 20°C. We can clearly see that increasing this threshold value gradually increases the difference in the Compton Edge value (Fig. 23). However, while for a threshold of 0.5°C, this deviation is still relatively small, it grows rapidly for larger values, exceeding 10% at 3°C. We conclude that introducing such a threshold into the temperature loop program is recommended and useful to prevent the loop from making small voltage corrections continuously, but this threshold value should not be too high. At a threshold value of 0.5°C, the difference in the Compton Edge value is less than 5%, so as a conclusion, we recommend that this parameter be set to a value equal to or less than 0.5°C.

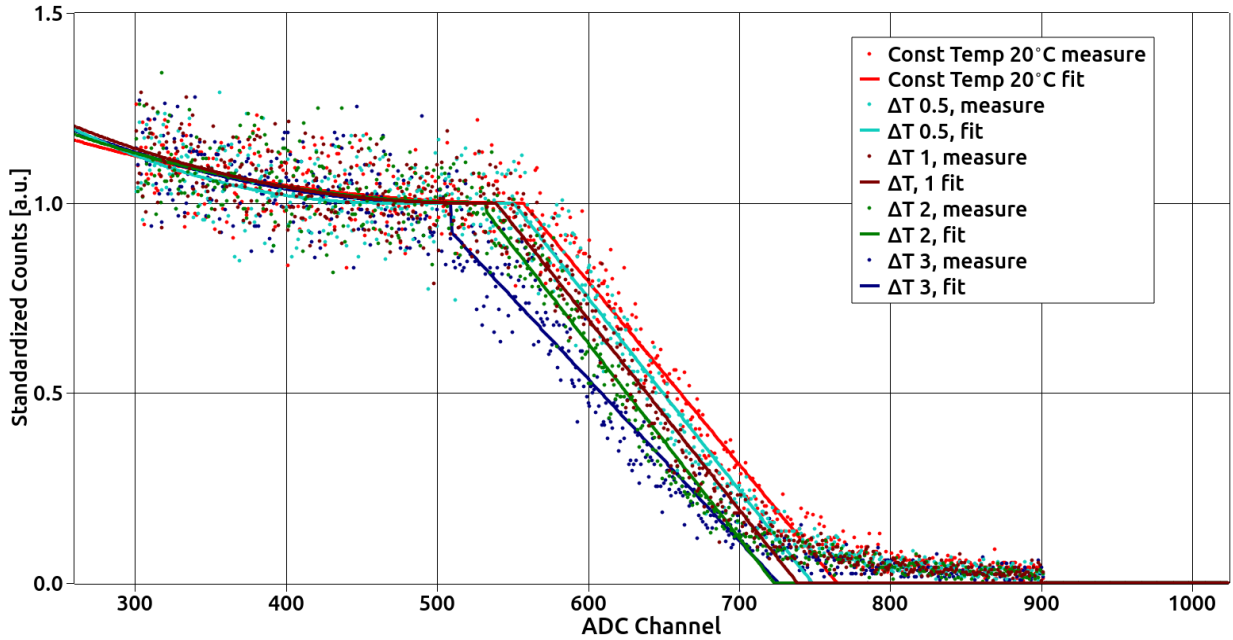

Figure 22. Comparison of the normalized spectrum graphs for measurements with an operating temperature loop, at different values of the threshold temperature (Dead Band), with the reference graph for the measurement with a constant temperature of 20°C. When the dead-band increases, the measured value of the Compton edge shifts towards lower values. Averaging was performed using a weighted exponential average with a coefficient of α=$10^{-5}$ and an averaging time of 0.1 s

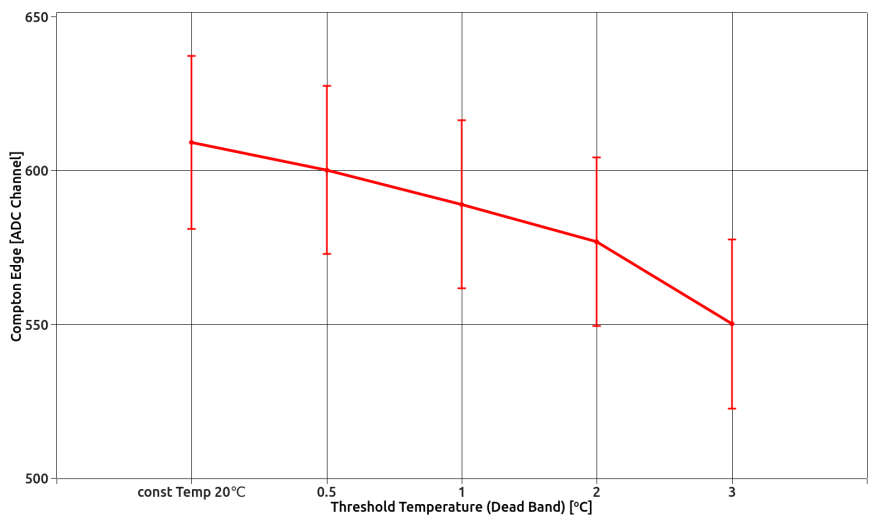

Figure 23. Dependence of the Compton edge value on the threshold temperature (Dead Band) parameter. When the dead-band increases, the measured value of the Compton edge shifts towards lower values

The next Figure 24 shows a series of measurements taken at different times during which the measured temperature values were collected and used to determine the average. The time values were 0.1, 1, and 10 seconds. We can clearly see that changing this time value does not noticeably affect the determined Compton Edge value (Fig. 25) and is not essential for the proper operation of our temperature loop under the tested conditions. This parameter may be important for very rapid temperature changes, which are not normaly encountered in laboratory or outdoor measurements.

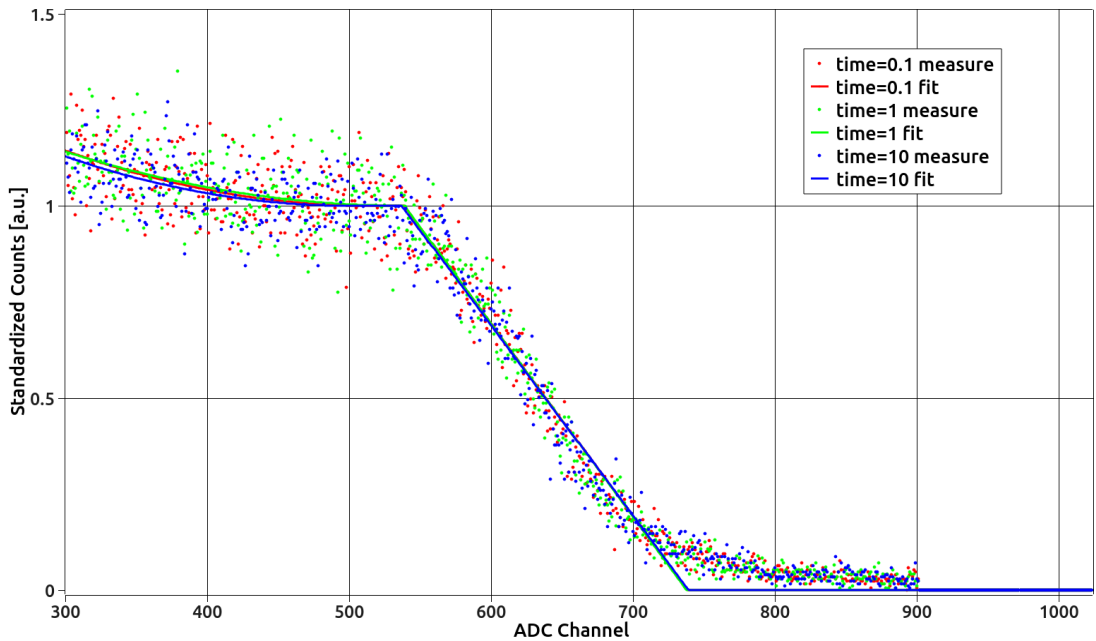

Figure 24. Comparison of temperature loop performance depending on different data collection times to calculate the average temperature (averaging times of 0.1 s, 1 s, and 10 s). The results within the measurement uncertainty limits coincide with each other. Averaging was performed using a weighted exponential average with a coefficient of $\alpha=10^{-5}$ and a temperature threshold step of 1°C.

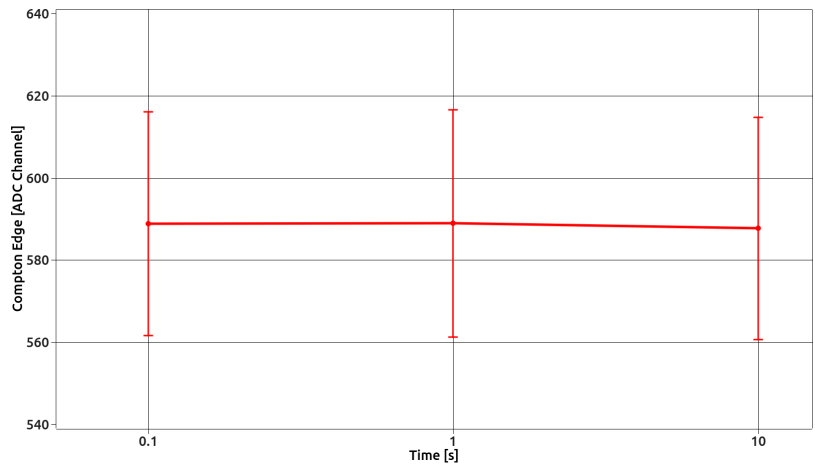

Figure 25. Comparison of Compton edge values as a function of different temperature data acquisition times used to calculate the average temperature.

The last series of measurements compared various methods for determining the average temperature value (Fig. 26). Many different average types were implemented as selectable options in the loop software (see Chapter 7). However, based on the simulations performed (Fig. 17), it can be concluded that using some of them yields very similar values. For this reason, comparative measurements were performed for only three of them, which yielded significantly different results in the simulations. We compared the performance of the simplest method (arithmetic mean) with two types of exponential weighted average with alpha coefficients of 1 and $10^{-5}$. From the obtained measurement results, it is clear that changing the average calculation method also does not noticeably affect the determined Compton Edge value (Fig. 27) and is not essential for the proper operation of our temperature loop under the tested conditions.

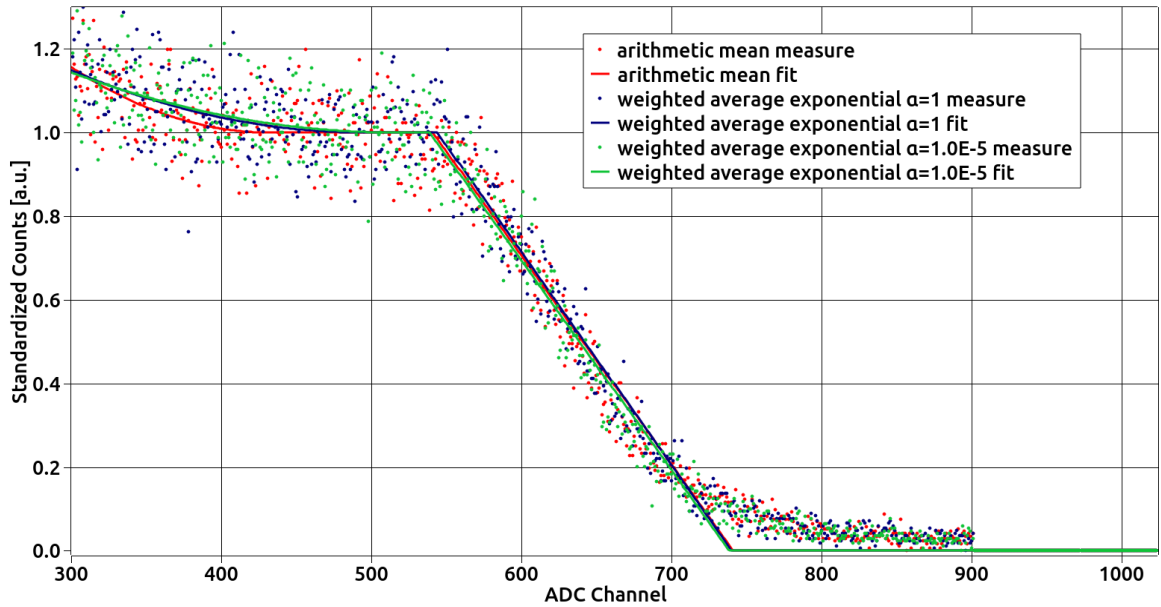

Figure 26. Comparison of the performance of the temperature loop using three different methods for calculating the average temperature value.: arithmetic mean, the exponentially weighted mean with an α coefficient of 1 (WE alpha 1), and the weighted mean with an α coefficient of $10^{-5}$. The results within the measurement uncertainty limits coincide with each other. Averaging was performed at a temperature threshold step of 1°C and an averaging time of 0.1 s

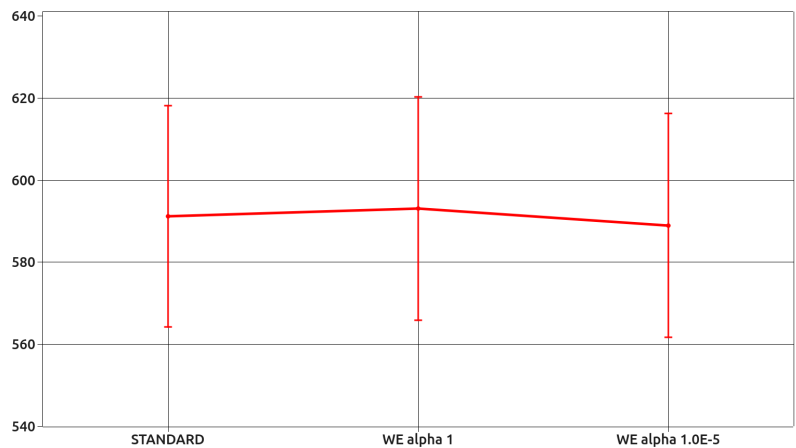

Figure 27. Dependence of the Compton edge values on the different temeperature averaging method

## 9. Summary

This article comprehensively presents all the issues related to the development of a properly and effectively operating temperature loop in detectors using semiconductor photomultipliers (SiPM) for light reading. The use of a loop is essential to maintaining a constant detector signal amplification level under changing ambient temperatures.

The design of the measurement setup, using a climatic chamber and small models of the actual detector, is described in detail, enabling reliable measurements. A fundamental parameter necessary for the operation of this type of detector is knowledge of the SiPM Temperature Coefficient Factor. We demonstrate a method for experimentally verifying this value provided by the SiPM sensor manufacturer and demonstrate that it is also necessary to take into account the influence of the electronics and the scintillator itself on the resulting value of this factor. In the case of our detector, this value changed by approximately 20%, from 50 mV to 62 mV, which was used in all subsequent measurements.

To enable effective comparison of temperature loop results, it was decided to use the Compton edge value in the studied spectrum. A detailed description of how to use and determine the Compton edge value in the studied spectra is provided. The method we propose is based on fitting (using Mathematica [12] procedures) the function (Eq. 3) to the measurement data and, based on the coefficients obtained from this fit, calculating the Compton edge value (Eq. 4). Importantly, the developed method works effectively even with low measurement statistics. An analysis of potential measurement errors related to the method used was also conducted.

Before the measurements began, a correction was made to the AFE electronics, significantly reducing the noise level and, consequently, the measurement uncertainty at extremely low currents. The AFE/HUB electronics control software was rebuilt and added temperature loop functionality, allowing for the modification of several key parameters of this loop's operation. The software was written using the Micropython environment for HUB and C utilizing the STM32F0 Hardware Abstraction Layer for AFE, and this article describes its basic features and the most important implemented variables.

Thanks to all this work, a series of laboratory measurements were successfully conducted demonstrating the effective operation of the designed temperature loop during multi-day measurements and the impact of changes in key parameters. These measurements demonstrated that changing the temperature sample collection time for calculating the average value, as well as the method of calculating the average value, are of no significant importance, and the user can change these parameters without compromising the effective and correct operation of the temperature loop. A crucial parameter is the temperature change threshold, after which the temperature loop automatically changes the SiPM supply voltage to maintain constant system gain. It was shown that introducing such a threshold is essential to ensure that the system does not change the supply voltage continuously and frequently. However, this threshold should not exceed 0.5°C to prevent the loop from negatively impacting the measurement system.

This article is a supplement to the previously published article [7], which described the operation of the detector electronics and their calibration for correct measurement performance. Therefore, both articles can be considered as a set.